\documentclass[manuscript]{aastex}

\slugcomment{Not to appear in Nonlearned J., 45.}

\shorttitle{Generalization of Levi-Civita regularization}
\shortauthors{Roman et al.}

\begin{document}


\title{Generalization of Levi-Civita regularization in the restricted three-body problem}


\author{R. Roman\altaffilmark{1} and I. Sz{\"u}cs-Csillik\altaffilmark{1}}
\affil{Astronomical Institute of Romanian Academy, Astronomical Observatory Cluj-Napoca, Str. Ciresilor No. 19, RO-400487 Cluj-Napoca, Romania }

\email{rdcroman@yahoo.com}

\email{iharka@gmail.com}

\begin{abstract}
A family of polynomial coupled function of $n$ degree  is proposed, in order to generalize the Levi-Civita regularization method, in the restricted three-body problem. Analytical relationship between polar radii in the physical plane and in the regularized plane are established; similar for polar angles. As a numerical application, trajectories of the test particle using polynomial functions of $2, 3,..., 8$ degree are obtained. For the polynomial of second degree, the Levi-Civita regularization method is found.
\end{abstract}


\section{Introduction}

The regularization (in Celestial Mechanics) is a
transformation of space and time variables, in order to eliminate
the singularities occurring in equations of motion. As Szebehely
show (see \cite{Sze1967}), the purpose of regularization is to
obtain regular differential equations of motion and not regular
solutions.

The regularization was introduced by Levi-Civita in 1906 (see \cite{Levi1906}) in plane, and generalized by Kustaanheimo and Stiefel in 1965 (see \cite{Kust1965}) in space. At the beginning, the regularization was developed for studying the singularities of Kepler motion, for analyzing the collisions of two point masses, and for improving the numerical integration of near-collision orbits.
Many studies of the regularization problem are in the restricted three-body problem, where there are two singularities. We can regularize local (one of them), or global. Birkhoff (1915), Thiele (1896), Burrau (1906), Lema\^{i}tre (1955), Arenstorf (1963), \'{E}rdi (2004), Sz\"{u}cs-Csillik and Roman (2012), and many other researchers studied the regularization of the restricted three-body problem.

In order to obtain the regularized equations of motion, one introduces a generating function \textit{S}, which depends on two harmonic and conjugated functions \textit{f} and \textit{g}. But there are many harmonic and conjugated functions. Using different couples of polynomial functions, one can obtain different methods of regularization. For the polynomial of second degree we obtain the Levi-Civita regularization method. By consequence, in this article we created a class of regularization methods, which all have in common the idea that \textit{f} and \textit{g} are harmonic and conjugate polynomial functions. We studied analytically some properties of these methods.

Starting from the graphical representation of the test particle's trajectory in the circular restricted three-body problem in the physical plane, and imposing a set of initial conditions, we obtained trajectories in the regularized plane, using 7 regularization methods. So, the methods can be compared not only by canonical equations of motion, but also by the shape of the obtained trajectories.

\section{The restricted three-body problem}

First of all, let us analyze why the regularization
is useful in the restricted three-body problem. For simplicity, we
shall consider in the following that the third body moves into the
orbital plane ($z=0$). Denoting $S_1$ and $S_2$ the components of
the binary system (whose masses are $m_1$ and $m_2$), the
equations of motion of the test particle (in the frame of the
restricted three-body problem) in the coordinate system
$(x,S_1,y)$, (the physical plane) are (see equations (1)-(2) in
\cite{RR2012}, \cite{RR2011}):
\begin{equation}\label{RR1}
\frac{d^2x}{dt^2}-2\frac{dy}{dt}=x-\frac{q}{1+q}-\frac{x}{(1+q)r_1^3}-\frac{q(x-1)}{(1+q)r_2^3}
\end{equation}

\begin{equation}\label{RR2}
\frac{d^2y}{dt^2}+2\frac{dx}{dt}=y-\frac{y}{(1+q)r_1^3}-\frac{q\:y}{(1+q)r_2^3}
\end{equation}
where
\begin{equation}\label{RR4}
r_1=\sqrt{x^2+y^2}\,\;,\;\;\;r_2=\sqrt{(x-1)^2+y^2}\;,\;\;\;\;\;q=\frac{m_2}{m_1}.
\end{equation}
These equations have singularities in terms $\frac{1}{r_1}$ and $\frac{1}{r_2}$ (see \cite{Csi2003}, \cite{Mioc2002}).
This situation corresponds to collision of the test particle with $S_1$ and $S_2$. If the test particle approaches very closely to one of the primaries,
such an event produces large gravitational force and sharp bends of orbit. Removing of these singularities can be done by regularization.

In order to regularize equations (1)-(2), we introduce the generalized coordinates $q_1$, $q_2$,  and the generalized momenta $p_1$, $p_2$ (see \cite{Boc1996} p. 266; \cite{RR2012}), and write the Hamiltonian and canonical equations of motion:
\begin{equation}
\mathcal{H}=\frac{1}{2}(p_1^2+p_2^2)+p_1 q_2- q_1 p_2 +\frac{q_1^2}{2}+\frac{q_2^2}{2}-\psi(q_1,q_2)\;.
\end{equation}
where
\begin{equation}
\psi(q_1,q_2)=\frac{1}{2}\left[\left(q_1-\frac{q}{1+q}   \right)^2+q_2^2+\frac{2}{(1+q)r_1}  +\frac{2q}{(1+q)r_2} \right]\;,
\end{equation}
 with
\begin{equation}
r_1=\sqrt{q_1^2+q_2^2},\;\;\;\;r_2=\sqrt{(q_1-1)^2+q_2^2}\;.
\end{equation}
Here the generalized coordinates and the generalized momenta were:
\begin{equation}
q_1=x\;, \;\;q_2=y\;,\;\;p_1=\dot{q_1}-q_2\;,\;\;\;p_2=\dot{q_2}+q_1\;.
\end{equation}
The canonical equations have the general form:
\begin{equation}
\dot {q_i}=\frac{\partial \mathcal{H}}{\partial p_i}\;,\;\;\;\dot {p_i}=-\frac{\partial \mathcal{H}}{\partial q_i}\;, \;\;\;\;i\in\{1,2 \}\;.
\end{equation}
The canonical equations obtained from equations (1)-(2) have, in the \\
$(q_1,S_1,q_2)$ coordinate system, the explicit form:
\begin{eqnarray}\label{eq1-3}
\frac{d q_1}{dt} &=& p_1+q_2 \\
\frac{d q_2}{dt} &=& p_2-q_1 \\
\frac{d p_1}{dt} &=& p_2 - \frac{q}{1+q}-\frac{1}{1+q}\cdot
\frac{q_1}{r_1^3} - \frac{q}{1+q} \cdot \frac{q_1-1}{r_2^3}  \\
\frac{d p_2}{dt} &=& -p_1 -\frac{1}{1+q}\cdot
\frac{q_2}{r_1^3} - \frac{q}{1+q} \cdot\frac{q_2}{r_2^3}.
\end{eqnarray}
The canonical equations have singularities in terms $\frac{1}{r_1}$ and $\frac{1}{r_2}$. These singularities can be eliminate by regularization.

\section{The regularization of the restricted three-body problem}
\label{sec:3}
The procedure of regularization consists on two transformations: \textit{the coordinate transformation}, which gives the shape of the orbit, and \textit{the time transformation}, which makes the slow-down motion.

\subsection{The coordinate transformation}

The first step performed in the process of coordinate transformation consists in introduction of new coordinates $Q_1$ and $Q_2$.
Let us introduce the generating function $\mathcal{S}$ (see \cite{Sti1971}, p.196):
\begin{equation}\label{eq1-7}
\mathcal{S}=-p_1 f(Q_1, Q_2)-p_2 g(Q_1, Q_2)\;,
\end{equation}
a twice continuously differentiable function. Stieffel (15) write that, in order to obtain the regularized equations of motion it is necessary to impose only one condition on functions $f$ and $g$: they must be harmonic and conjugated function, that means we have: 
\begin{eqnarray}
\frac{\partial f}{\partial Q_1} &=& \frac{\partial g}{\partial Q_2}\nonumber\\
\frac{\partial f}{\partial Q_2} &=& -\frac{\partial g}{\partial Q_1}\;\;.\nonumber
\end{eqnarray}
The generating equations are
\begin{equation}
q_i = -\frac{\partial \mathcal{S}}{\partial p_i}, \;\;\;\;P_i =-\frac{\partial \mathcal{S}}{\partial Q_i}\;,\;\;\;\;i\in \{1, 2 \}\;,
\end{equation}
with $P_1,\;P_2$ as new generalized momenta, or explicitly
\begin{eqnarray}\label{eq1-9}
q_1 &=& -\frac{\partial \mathcal{S}}{\partial p_1} =f(Q_1, Q_2) \nonumber\\
q_2 &=& -\frac{\partial \mathcal{S}}{\partial p_2}=g(Q_1, Q_2) \nonumber\\
P_1 &=&- \frac{\partial \mathcal{S}}{\partial Q_1}=p_1 \frac{\partial f}{\partial Q_1}+p_2 \frac{\partial g}{\partial Q_1}=p_1 a_{11}+p_2 a_{12} \nonumber\\
P_2 &=& -\frac{\partial \mathcal{S}}{\partial Q_2}=p_1 \frac{\partial f}{\partial Q_2}+p_2 \frac{\partial g}{\partial Q_2}=-p_1 a_{12}+p_2 a_{11}
\end{eqnarray}
where
\begin{eqnarray}
a_{11}&=&\frac{\partial f}{\partial Q_1} = \frac{\partial g}{\partial Q_2}\nonumber\\
a_{12}&=&-\frac{\partial f}{\partial Q_2} = \frac{\partial g}{\partial Q_1}\nonumber
\end{eqnarray}
Let introduce the following notation:
$${\bf A}= \left( \matrix{a_{11} & a_{12} \cr -a_{12} & a_{11}} \right), \;\; D=det {\bf A}= a_{11}^2+a_{12}^2\;,$$
\begin{equation}
{\bf p}=\left( \matrix{p_1 \cr p_2 } \right),\;{\bf P}=\left( \matrix{P_1 \cr P_2} \right), \;\textbf{p}=\frac{A^T}{D}\textbf{P},\;{\bf P}=A\cdot  {\bf p},\;p_1^2+p_2^2=\frac{P_1^2+P_2^2}{D}
\end{equation}
where $\textbf{A}^T$ represents the transpose of matrix $\textbf{A}$.

The new Hamiltonian with the generalized coordinates $Q_1$ and $Q_2$ and generalized momenta $P_1$ and $P_2$ is:
\begin{eqnarray}\label{eq1-11}
\mathcal{H}(Q_1,Q_2,P_1,P_2) &=& \frac{1}{2D} \left[ P_1^2+P_2^2 +P_1 \frac{\partial }{\partial Q_2} (f^2+g^2) - P_2 \frac{\partial }{\partial Q_1} (f^2+g^2) \right] +  \nonumber \\
 &+& \frac{q}{1+q}f- \frac{1}{1+q}\cdot \frac{1}{\overline{r}_1} - \frac{q}{1+q} \cdot \frac{1}{\overline{r}_2}-\frac{q^2}{2(1+q)^2}
\end{eqnarray}
where $\overline{r}_1=\sqrt{f^2+g^2}$,
$\overline{r}_2=\sqrt{(f-1)^2+g^2}$, $D=a_{11}^2+a_{12}^2$ and the
explicit canonical equations of motion in new variables become:
\begin{eqnarray}\label{eq1-12}
\frac{dQ_1}{dt}&=&\frac{1}{2D} \left[ 2P_1+\frac{\partial }{\partial Q_2} (f^2+g^2) \right] \\
\frac{dQ_2}{dt}&=&\frac{1}{2D} \left[ 2P_2-\frac{\partial }{\partial Q_1} (f^2+g^2) \right ] \nonumber\\
\frac{dP_1}{dt}&=&- \frac{P_1}{2D} \cdot \frac{\partial^2 }{\partial Q_1\partial Q_2} (f^2+g^2) +  \frac{P_2}{2D} \cdot \frac{\partial ^2}{\partial Q_1 \partial Q_1} (f^2+g^2) - \frac{q}{1+q} \frac{\partial f}{\partial Q_1}+ \nonumber \\
&+&  \frac{1}{1+q}\cdot \frac{\partial}{ \partial Q_1} \left( \frac{1}{\overline{r}_1} \right) + \frac{q}{1+q} \cdot \frac{\partial}{ \partial Q_1} \left( \frac{1}{\overline{r}_2}  \right) \nonumber \\
\frac{dP_2}{dt}&=&- \frac{P_1}{2D} \cdot \frac{\partial^2 }{\partial Q_2\partial Q_2} (f^2+g^2) +  \frac{P_2}{2D} \cdot \frac{\partial ^2}{\partial Q_2 \partial Q_1} (f^2+g^2) - \frac{q}{1+q} \frac{\partial f}{\partial Q_2}+ \nonumber \\
&+&  \frac{1}{1+q}\cdot \frac{\partial}{ \partial Q_2} \left( \frac{1}{\overline{r}_1} \right) + \frac{q}{1+q} \cdot \frac{\partial}{ \partial Q_2} \left( \frac{1}{\overline{r}_2}  \right)\nonumber
\end{eqnarray}

So, in order to obtain the regularized equations of motion (18), it was necessary to impose only one condition on functions $f$ and $g$: they must be harmonic conjugated functions. But there are a lot of harmonic conjugated functions. 

We can find harmonic and conjugate polynomial functions, by using the theory of complex functions. We denote $z=Q_1+iQ_2$ a single complex variable and $h:\Omega \rightarrow \mathbb{C}$, $h(z)=h(Q_1+iQ_2)=f(Q_1,Q_2)+ig(Q_1,Q_2)$ a complex-valued function. (Here $f$ and $g$ are two real functions, depending on two real variables $Q_1$ and $Q_2$.) From the theory of complex numbers (see \cite{Cara:2001}) we know that, if $h(z)$ is a complex function, then its real and imaginary parts are harmonic functions. That means:

$\frac{\partial^2f(Q_1,\;Q_2)}{\partial Q_1^2}+\frac{\partial^2f(Q_1,\;Q_2)}{\partial Q_2^2}=0$, and
$\frac{\partial^2g(Q_1,\;Q_2)}{\partial Q_1^2}+\frac{\partial^2g(Q_1,\;Q_2)}{\partial Q_2^2}=0$.

Considering $h(z)=z$ it results $h(z^n)=z^n$, $n\in \mathbb{N}$ and $z^n=(Q_1+iQ_2)^n$. We obtain so the harmonic polynomials (see Table 1, for $n=0, 1, ..., 8$).

\begin{table}
\begin{center}
    \caption{Some harmonic and conjugate polynomial functions}
    \label{tab:1}
        \begin{tabular}{cll}
             \hline\noalign{\smallskip}
             $n$ & $f(Q_1,Q_2$) & $g(Q_1,Q_2)$  \\
             \noalign{\smallskip}\hline\noalign{\smallskip}
             $n=0$ & $1$ & $0$ \\
             $n=1$ & $Q_1$ & $Q_2$ \\
             $n=2$ & $Q_1^2-Q_2^2$ & $2Q_1Q_2$ \\
             $n=3$ & $Q_1^3-3Q_1Q_2^2$ & $3Q_1^2Q_2-Q_2^3$ \\
             $n=4$ & $Q_1^4-6Q_1^2Q_2^2+Q_2^4$ & $4Q_1^3Q_2-4Q_1Q_2^3$  \\
             $n=5$ & $Q_1^5-10Q_1^3Q_2^2+5Q_1Q_2^4$ & $5Q_1^4Q_2-10Q_1^2Q_2^3+Q_2^5$  \\
             $n=6$ & $Q_1^6-15Q_1^4Q_2^2+15Q_1^2Q_2^4-Q_2^6$ & $6Q_1^5Q_2-20Q_1^3Q_2^3+6Q_1Q_2^5$  \\
             $n=7$ & $Q_1^7-21Q_1^5Q_2^2+35Q_1^3Q_2^4-7Q_1Q_2^6$ & $7Q_1^6Q_2-35Q_1^4Q_2^3+21Q_1^2Q_2^5-Q_2^7$  \\
             $n=8$ & $Q1^8-28Q1^6Q2^2+70Q1^4Q2^4-$ & $8Q1^7Q2-56Q1^5Q2^3+56Q1^3Q2^5$-\\
 & $-28Q1^2Q2^6+Q2^8$ & $-8Q1Q2^7$ \\
             \noalign{\smallskip}\hline
        \end{tabular}
\end{center}
\end{table}

\textit{Remarks}:
\begin{enumerate}
    \item The case $n=0$ (in Table 1) isn't relevant for regularization.
    \item If $n=1$ (in Table 1), the physical plane coincides with  the regularized plane.
    \item If $n=2$ (in Table 1) , we obtain the Levi-Civita regularization (see  \citep{Levi1906}).
    \item It is easy to see that all pairs of functions $f(Q_1, Q_2)$ and $g(Q_1, Q_2)$ are conjugate, because they verify the relations of Cauchy-Riemann:

$\frac{\partial f}{\partial Q_1}=\frac{\partial g}{\partial Q_2}$ and $\frac{\partial f}{\partial Q_2}=-\frac{\partial g}{\partial Q_1}$.

\end{enumerate}

In order to obtain the canonical equations, when $f$ and $g$ are harmonic and conjugate polynomial functions, we have to write first the corresponding Hamiltonian equation.

Let us consider the complex variable $z=Q_1+i\; Q_2$, which can be
written in the trigonometric form:
$z=\sqrt{Q_1^2+Q_2^2}\;(\cos(T)+i\;\sin(T))$, where
$\tan(T)=\frac{Q_2}{Q_1}$, $i=\sqrt{-1}$. We denote $f_n=\Re(z^n)$ and $g_n=\Im(z^n)$, $n\in \;\mathbb{N}$, $n\geq2$. By consequence we
obtain:
$$ f_n^2+g_n^2=(Q_1^2+Q_2^2)^n\;,\;\;\;D_n=\left(\frac{\partial f_n}{\partial Q_1}\right)^2+\left(\frac{\partial f_n}{\partial Q_2}\right)^2=n^2\;(Q_1^2+Q_2^2)^{n-1} \;\;. $$
The equation of Hamiltonian (eq. 17) becomes in this case:
\begin{eqnarray}\label{HPolnD}
\mathcal{H}(Q_1, Q_2,P_1, P_2) &=& \frac{1}{2D_n} (P_1^2+P_2^2) + \frac{P_1 Q_2}{n} -\frac{P_2Q_1}{n}+\frac{q}{1+q}f_n-\nonumber\\
&-& \frac{1}{1+q}\cdot \frac{1}{\overline{r}_{1n}} - \frac{q}{1+q} \cdot \frac{1}{\overline{r}_{2n}}-\frac{q^2}{2(1+q)^2}
\end{eqnarray}

and the canonical equations of motion will be:
\begin{eqnarray}\label{eqPolnD}
\frac{dQ_1}{dt}&=&  \frac{P_1}{D_n}+\frac{Q_2}{n}\nonumber\\
\frac{dQ_2}{dt}&=&  \frac{P_2}{D_n}-\frac{Q_1}{n}\nonumber\\
\frac{dP_1}{dt}&=& \frac{(n-1)Q_1(P_1^2+P_2^2)}{D_n(Q_1^2+Q_2^2)}+\frac{P_2}{n}
-\frac{q}{1+q}\frac{\partial f_n}{\partial Q_1} +\nonumber\\
&+& \frac{1}{(1+q)} \frac{\partial}{\partial Q_1} \left( \frac{1}{\overline{r}_{1n}}\right) +\frac{q}{(1+q)} \frac{\partial }{\partial Q_1} \left( \frac{1}{\overline{r}_{2n}} \right) \\
\frac{dP_2}{dt} &=& \frac{(n-1)Q_2(P_1^2+P_2^2)}{D_n(Q_1^2+Q_2^2)}-\frac{P_1}{n}
-\frac{q}{1+q} \frac{\partial f_n}{\partial Q_2} +\nonumber\\
&+& \frac{1}{(1+q)}\frac{\partial }{\partial Q_2} \left( \frac{1}{\overline{r}_{1n}} \right) + \frac{q}{(1+q)}\frac{\partial }{\partial Q_2} \left(\frac{1}{\overline{r}_{2n}} \right) \nonumber
\end{eqnarray}

where $\overline{r}_{1n}=\sqrt{f_n^2+g_n^2}\;,\;\;\;\overline{r}_{2n}=\sqrt{(f_n-1)^2+g_n^2}$.

As one can see these equations are still singular. In order to eliminate the singularities we must transform the time.

\subsection{The time transformation}

The second step performed in the process of regularization consists in the time transformation see \citep{Sti1971}.
In order to solve the canonical equations of motion, we introduce the fictitious time $\tau$, and make the time transformation
$\frac{dt}{d\tau}=\overline r_{1n}^3 \overline r_{2n}^3$. So, the regular equations of motion become:

\begin{eqnarray}
\frac{dQ_1}{d\tau}&=& \left( \frac{P_1}{D_n}+\frac{Q_2}{n}\right)\overline r_{1n}^3 \overline r_{2n}^3\nonumber\\
\frac{dQ_2}{d\tau}&=& \left( \frac{P_2}{D_n}-\frac{Q_1}{n}\right)\overline r_{1n}^3 \overline r_{2n}^3\nonumber\\
\frac{dP_1}{d\tau}&=&( \frac{(n-1)Q_1(P_1^2+P_2^2)}{D_n(Q_1^2+Q_2^2)}+\frac{P_2}{n}
-\frac{q}{1+q}\frac{\partial f_n}{\partial Q_1} +\nonumber\\
&+& \frac{1}{(1+q)} \frac{\partial}{\partial Q_1} \left( \frac{1}{\overline{r}_{1n}}\right) +\frac{q}{(1+q)} \frac{\partial }{\partial Q_1} \left( \frac{1}{\overline{r}_{2n}} \right) )\overline r_{1n}^3 \overline r_{2n}^3  \\
\frac{dP_2}{d\tau} &=& (\frac{(n-1)Q_2(P_1^2+P_2^2)}{D_n(Q_1^2+Q_2^2)}-\frac{P_1}{n}
-\frac{q}{1+q} \frac{\partial f_n}{\partial Q_2} +\nonumber\\
&+& \frac{1}{(1+q)}\frac{\partial }{\partial Q_2} \left( \frac{1}{\overline{r}_{1n}} \right) + \frac{q}{(1+q)}\frac{\partial }{\partial Q_2} \left(\frac{1}{\overline{r}_{2n}} \right))\overline r_{1n}^3 \overline r_{2n}^3 \nonumber
\end{eqnarray}
Now, the equations of motion of the test particle do not have singularities.

\section{Numerical application: trajectories of the test particle}

In order to obtain the trajectories in the regularized plane for different polynomial functions $f$ and $g$, the canonical equations of motion of the test particle must be integrated, using initial conditions. We denote
$$  q_{10}=q_1(t)/_{_{t=0}}\;,\;\;q_{20}=q_2(t)/_{_{t=0}}\;,\;\;p_{10}=p_1(t)/_{_{t=0}}\;,\;\;p_{20}=p_2(t)/_{_{t=0}}$$
the initial conditions for the canonical equations (9), (10), (11), (12) in the physical plane, and
$$  Q_{10}=Q_1(t)/_{_{t=0}}\;,\;\;Q_{20}=Q_2(t)/_{_{t=0}}\;,\;\;P_{10}=P_1(t)/_{_{t=0}}\;,\;\;P_{20}=P_2(t)/_{_{t=0}}$$
the initial conditions for the canonical equations in the regularized plane. The connection between these initial conditions is given by the equations (see \cite{RR2012}):
\begin{eqnarray}
q_{10}&=&f(Q_{10},Q_{20})\nonumber\\
q_{20}&=&g(Q_{10},Q_{20})\nonumber\\
P_{10}&=&p_{10}\left(\frac{\partial f}{\partial Q_{1}}\right)_{(Q_{10},\;Q_{20})}+p_{20}\left(\frac{\partial g}{\partial Q_{1}}\right)_{(Q_{10},\;Q_{20})}\\
P_{20}&=&-p_{10}\left(\frac{\partial g}{\partial Q_{1}}\right)_{(Q_{10},\;Q_{20})}+p_{20}\left(\frac{\partial f}{\partial Q_{1}} \right) _{(Q_{10},\;Q_{20})}  \nonumber
\end{eqnarray}

Applying the transformation of coordinates given by $f$ and $g$, not only the trajectory will be changed, but the positions of the components of the binary system $S_1$ and $S_2$ will be changed too. In Table 2 there are given the positions of $S_1$ and $S_2$ in the regularized plane, for every type of transformation presented in Table 1. Sometimes to one position of $S_2$ in the physical plane correspond two or more positions in the regularized plane. This fact is normal (see \cite {Sti1971}), and is shown in Table 2.

Starting with the initial conditions in the physical plane:
$$ q_{10}=0.6\;,\;\;q_{20}=0.4\;,\;\;p_{10}=0.1\;,\;\;p_{20}=0.6\;, $$
we calculated the corresponding initial conditions in the regularized plane, for each regularization method presented in Table 1. These new initial conditions are given in Table 2, as well.

\begin{table}
\begin{center}
    \caption{$S_1$, $S_2$ positions and the initial conditions for different methods of regularization (the case $n=1$ corresponds to the physical plane)}
    \label{tab:3}
      \begin{minipage}{400pt}
        \begin{tabular}{llcc}
             \hline\noalign{\smallskip}
             Type &  $S_1$ & $S_2$\footnote{where $S_2^k\;(\cos\frac{2\pi(k-1)}{n}, \;\sin\frac{2\pi(k-1)}{n} )$, $\;\;k\in\{ 1,...,n \}$ }&$(Q_{10},Q_{20},P_{10},P_{20})$\footnote{ calculated only for $k=1$  } \\
             \noalign{\smallskip}\hline\noalign{\smallskip}
Pol. function, $n=1$ &  $(0,0)$ & $(1,0)$    & $(0.6, 0.4, 0.1, 0.6)$         \\
Pol. function, $n=2$ &  $(0,0)$ & $S_2^k , k\in\{ 1,2  \}$    & $(0.812, 0.246, 0.457, 0.926)$ \\
Pol. function, $n=3$ &  $(0,0)$ & $S_2^k , k\in\{ 1,...,3  \}$    & $(0.879, 0.174, 0.775, 1.245)$ \\
Pol. function, $n=4$ &  $(0,0)$ & $S_2^k , k\in\{ 1,...,4  \}$    & $(0.911, 0.135, 1.084, 1.564)$ \\
Pol. function, $n=5$ &  $(0,0)$ & $S_2^k , k\in\{ 1,...,5  \}$    & $(0.930, 0.109, 1.389, 1.884)$ \\
Pol. function, $n=6$ &  $(0,0)$ & $S_2^k , k\in\{ 1,...,6  \}$    & $(0.942, 0.092, 1.693, 2.203)$ \\
Pol. function, $n=7$ &  $(0,0)$ & $S_2^k , k\in\{ 1,...,7  \}$    & $(0.950, 0.080, 1.995, 2.523)$ \\
Pol. function, $n=8$ &  $(0,0)$ & $S_2^k , k\in\{ 1,...,8  \}$    & $(0.957, 0.070, 2.297, 2.843)$ \\
             \noalign{\smallskip}\hline
        \end{tabular}
    \end{minipage}    
\end{center}
\end{table}

In Figure 1 there are presented the trajectories for each type of functions $f$ and $g$, given in Table 1. All the trajectories are represented for a time interval equal with an orbital period. For $n=1$ is represented the trajectory of the test particle in the physical plane. For $n=2$, the trajectory in the Levi-Civita regularized plane is given.

\begin{figure}
\begin{center}
  \includegraphics[height=0.8\textheight]{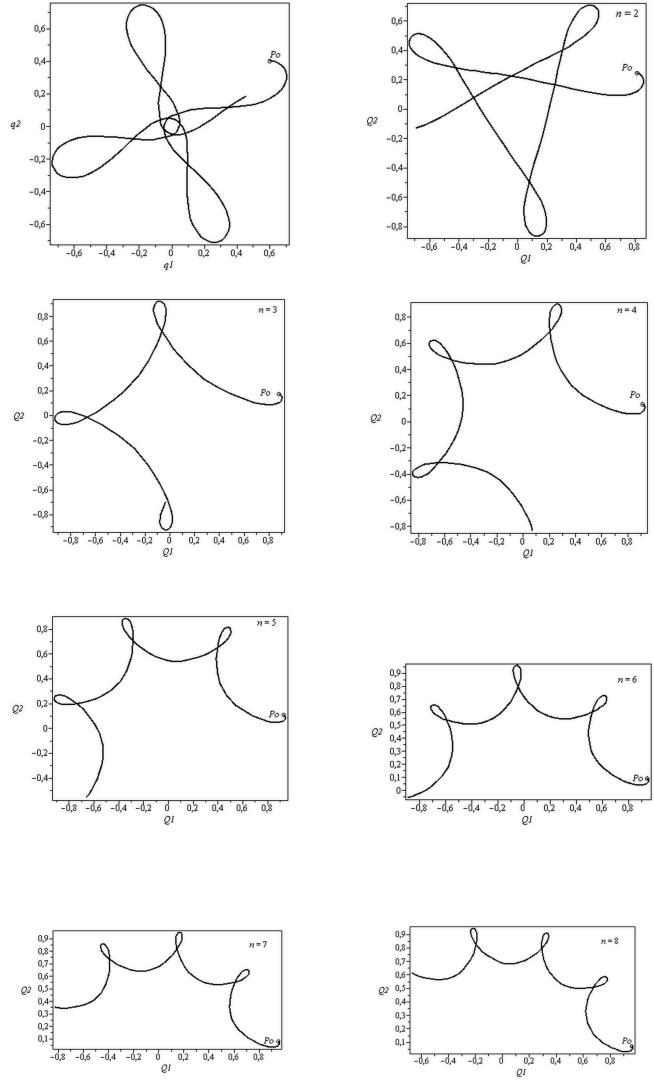}
  \caption{Trajectories of the test particle, when $f$ and $g$ are polynomial functions of order $n=2...8$}
  \end{center}
\end{figure}

As we can see from Figure 1, by changing the degree $n$ of polynomial functions $f$ and $g$, the initial position $P_0$ is changed, but the shape of the trajectory maintains the same topology.

If $f$ and $g$ are polynomial functions of $n$ degree, the greater
is $n$, the greater is the distance of $P_0$ to the origin of the
coordinate system.

In \cite{RR2012} there are presented some considerations on the geometrical transformation for Levi-Civita regularization. In the following we generalize two results concerning the polar radii and the polar angles for each values of polynomial degrees (see Figure 2).

\begin{figure}[http]
\begin{center}
   \includegraphics[height=0.2\textheight]{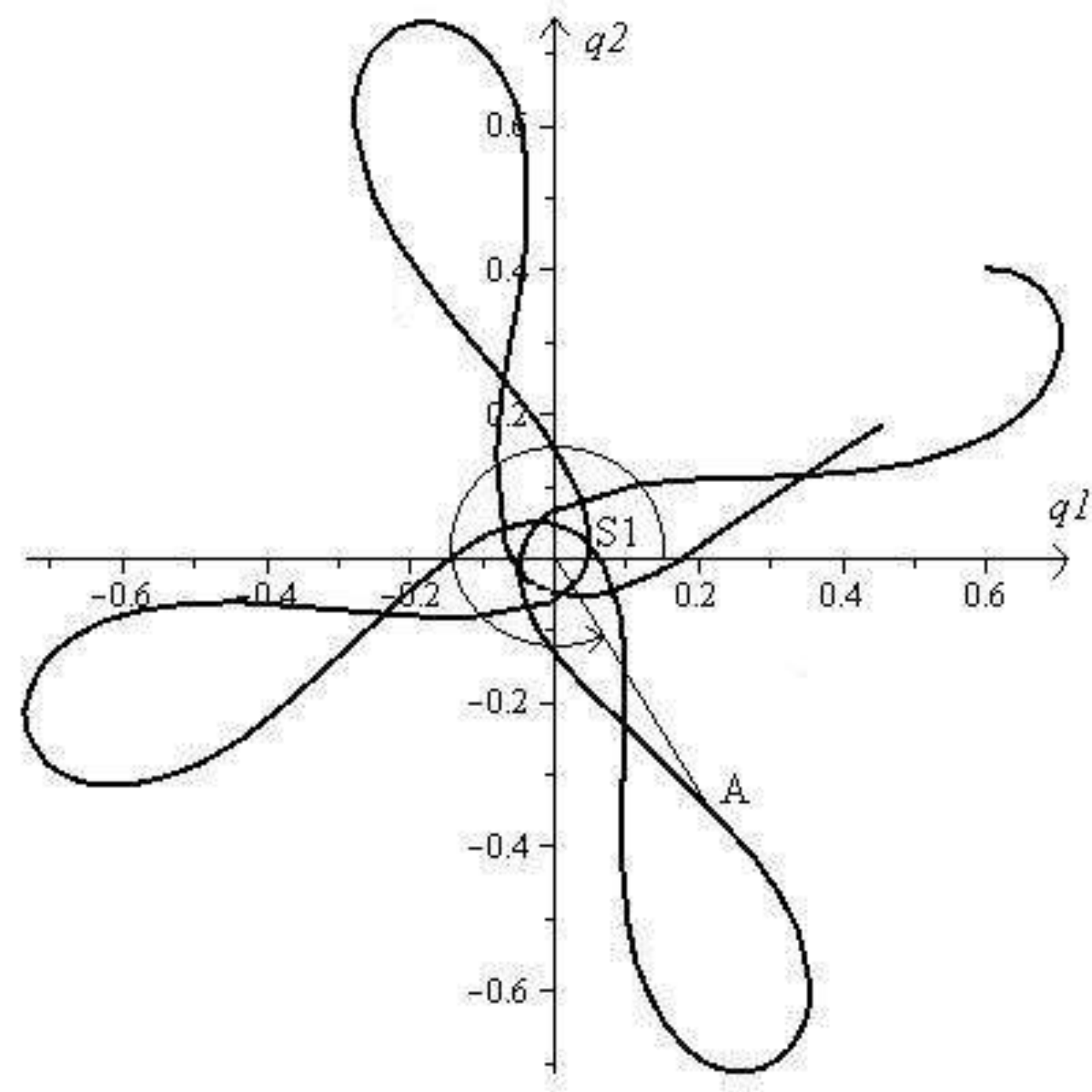}
   \includegraphics[height=0.2\textheight]{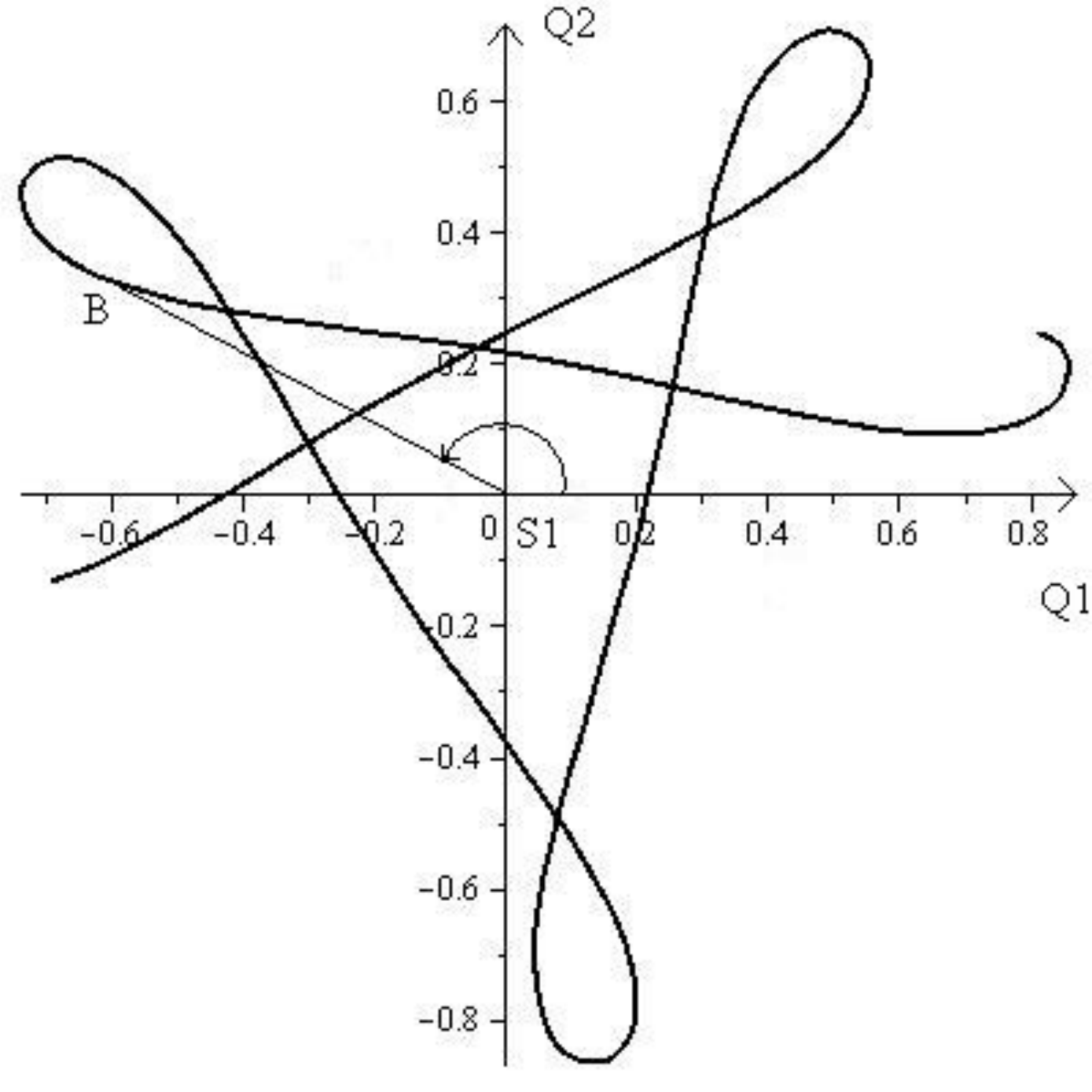}
  \caption{How to obtain the B point in (Q1, S1, Q2) plane from the A point in (q1, S1, q2) plane, for $n=2$}
  \end{center}
\end{figure}

\textbf{Theorem}

\textit{In the polynomial regularization's methods, if A is an arbitrary point of the trajectory in the physical plane ($n=1$), and B is its corresponding point in the regularized plane (see Figure 2), then we have the following relations concerning the polar radii and angles}:
$$|S_1A|=|S_1B|^n\;,\;\;\;\;  \widehat{AS_1q_1}=n\cdot \widehat{BS_1Q_1}\;,\;\;\;\; \forall\;n\in \mathbb{N},\;\;n\geq1.$$\\
\textit{Proof}:\\
We have:
$$|S_1A|=\sqrt{q_1^2+q_2^2},\;\;|S_1B|=\sqrt{Q_1^2+Q_2^2},$$
$$q_1=f_n=(\sqrt{Q_1^2+Q_2^2})^n\;\cos(nT)\;,\;\;q_2=g_n=(\sqrt{Q_1^2+Q_2^2})^n\;\sin(nT)$$
Then
\begin{eqnarray}
|S_1A|&=&\sqrt{q_1^2+q_2^2}= (\sqrt{Q_1^2+Q_2^2})^n(\sin^2(nT)+\cos^2(nT))= \nonumber\\
&=&(\sqrt{Q_1^2+Q_2^2})^n = |S_1B|^n \;.\nonumber
\end{eqnarray}
On the other hand
\begin{eqnarray}
\tan(\widehat{AS_1q_1})&=&\frac{q_2}{q_1}=\frac{g_n}{f_n}=\tan(nT)=\tan(n\;\arctan(\frac{Q_2}{Q_1}))= \nonumber\\
&=&\tan(n\;\widehat{BS_1Q_1}) , \nonumber
\end{eqnarray}
and then: $\widehat{AS_1q_1}=n\cdot\widehat{BS_1Q_1}+k\pi, \;\; k\in \mathbb{Z}$, or 
\begin{equation}\label{eq23}
\widehat{BS_1Q_1}=\frac{1}{n}\;\widehat{AS_1q_1}-\frac{k\pi}{n},\;\;k\in \mathbb{Z}.
\end{equation}
From the notation $f_n=\Re(z^n)$ and $g_n=\Im(z^n)$, $n\in \; \mathbb{N}$, $n\geq2$ we can see that to one point in the regularized plane (one couple $(Q_1,Q_2)$) it corresponds only one point in the physical plane (one couple $(q_1,q_2)$), then in equation \ref{eq23} $k$ must have only one value, for all values of $n$. If $n=1$ the physical plane is transformed in physical plane (the identical transformation), and for this transformation $\widehat{BS_1Q_1}=\widehat{AS_1q_1}$. It results that the unique value of $k$ must be $k=0$. Then, $\widehat{BS_1Q_1}=\widehat{AS_1q_1},  \forall\;n\in \mathbb{N},\;\;n\geq1.$

If $n=2$ one can obtain the Levi-Civita's result.

\section{Conclusion}

Because the unique condition imposed to the generating functions \textit{f} and \textit{g} is to be harmonic and conjugate, it exists a whole family of polynomial functions which can be selected. These polynomial functions of \textit{n} degree generate trajectories having the same topology. The greater is the degree of the polynomial, the greater is the distance of initial start point to the origin of the coordinate system. 

The polar radius in the physical plane is equal whith the polar radius in the regularized plane, raised to the \textit{n} power.

The polar angle in the physical plane is \textit{n} times greater than the polar angle in the regularized plane.

For \textit{n}=2 the Levi-Civita regularization methods is found. 

Even if Levi-Civita regularization method is more easy to use, from theoretical point of view it is interesting to encapsulate this method in a family of methods which all conserve the Levi-Civita method's properties.

\acknowledgments

The authors wish to acknowledge the anonymous reviewer for his/her helpful comment to the manuscript.


\begin{thebibliography}{}
\bibitem[Arenstorf, 1963]{Are1963} Arenstorf, R. F.: AJ {\bf 68}, 548 (1963)
\bibitem[Birkhoff, 1915]{Birk1915} Birkhoff, G. D.: {Rend. Circ. Mat. Palermo} {\bf 39}, 1 (1915)
\bibitem[Boccaletti and Pucacco, 1996]{Boc1996} Boccaletti, D., Pucacco, G.:  {Theory of orbits} {\bf 1}. Springer-Verlag Berlin Heidelberg New York (1996)
\bibitem[Burrau, 1906]{Burrau1906} Burrau, C.: {Vierteljahrschrift Astron. Ges.} {\bf 41}, 261 (1906)
\bibitem[Carath\'{e}odory, 2001]{Cara:2001} Carath\'{e}odory: Theory of functions of a complex variable, Vol. 1, 304 pages, AMS Chelsea Publishing, Providence, Rhode Island (2001)
\bibitem[Csillik, 2003]{Csi2003} Csillik, I.: Regularization methods in celestial mechanics, Cluj: House of the Book of Science (2003)
\bibitem[\'{E}rdi, 2004]{Erdi2004} \'{E}rdi, B.: {Celestial Mechanics} {\bf 90}, 35 (2004)
\bibitem[Kustaanheimo and Stiefel, 1965]{Kust1965} Kustaanheimo, P., Stiefel, E. L.: {J. Reine Angewandte Math.} {\bf 218}, 204 (1965)
\bibitem[Lema\^{i}tre, 1955]{Lema1955} Lema\^{i}tre, G.: {Vistas in Astronomy} {\bf 1}, 207 (1955)
\bibitem[Levi-Civita, 1906]{Levi1906} Levi-Civita, T.: {Acta Mathematica} {\bf 30}, 305 (1906)
\bibitem[Mioc and Csillik, 2002]{Mioc2002} Mioc, V., Csillik, I.: RoAJ {\bf 12}, 167 (2002) 
\bibitem[Roman, 2011]{RR2011} Roman, R., Ap\&SS {\bf 335}, 475 (2011) 
\bibitem[Roman et al., 2012]{RR2012} Roman, R., Sz\"{u}cs-Csillik, I.:  Ap\&SS {\bf 338}, 233 (2012)
\bibitem[Sz\"{u}cs-Csillik and Roman, 2012]{CZRR2012} Sz\"{u}cs-Csillik, I., Roman, R.: RoAJ {\bf 22}, 2, 145 (2012) 
\bibitem[Stiefel and Scheifele, 1971]{Sti1971} Stiefel, L., Scheifele, G.: {Linear and regular celestial mechanics}. Springer Berlin (1971)
\bibitem[Szebehely, 1967]{Sze1967} Szebehely, V.: Theory of orbits. Academic Press, New York (1967) 
\bibitem[Thiele, 1896]{Thiele1896} Thiele, T. N.: {Astron. Nachr.} {\bf 138}, 1 (1896)
\end{thebibliography}
\end{document}